# Simplified Aluminum Nitride Processing for Low-Loss Integrated Photonics and Nonlinear Optics

Haochen Yan[1,2,+], Shuangyou Zhang[3,1*,+], Arghadeep Pal[1,2], Alekhya Gosh[1,2], Abdullah Alabbadi[1,2], Masoud Kheyri[1,2], Toby Bi[1,2], Yaojing Zhang[1,4], Irina Harder[1], Olga Lohse[1], Florentina Gannott[1], Alexander Gumann[1], Eduard Butzen[1], Katrin Ludwig[1], and Pascal Del'Haye[1,2,*]

[1]*Max Planck Institute for the Science of Light, 91058 Erlangen, Germany*

[2]*Department of Physics, Friedrich-Alexander-Universität Erlangen-Nürnberg, 91058 Erlangen, Germany*

[3]*Department of Electrical and Photonics Engineering, Technical University of Denmark, Kgs. Lyngby, 2800, Denmark*

[4]*School of Science and Engineering, The Chinese University of Hong Kong, Shenzhen, Guangdong 518172, China*

+ These authors contribute equally

* Corresponding authors: shzhan@dtu.dk, pascal.delhaye@mpl.mpg.de

## Abstract

Aluminum nitride (AlN) is an extremely promising material for integrated photonics because of the combination of strong $\chi^{(2)}$ and $\chi^{(3)}$ nonlinearities. However, the intrinsic hardness of the material and charging effects during electron beam lithography make AlN nanofabrication a challenging process. Conventional approaches often require multiple hard masks and a metal mask to fabricate nanostructures. In this letter, we report a novel, simple method to fabricate AlN microresonators by using a single layer of silicon nitride mask combined with a thin conductive polymer layer. The conductive layer can be conveniently removed during developing without requiring an additional etching step. We achieve high intrinsic quality ($Q$) factors up to $1.0 \times 10^6$ in AlN microresonators and demonstrate several nonlinear phenomena within our devices, including frequency comb generation, Raman lasing, third harmonic generation and supercontinuum generation.

# I. Introduction

Aluminum nitride (AlN) has gained significant attention in the field of integrated photonics for its unique properties including a large bandgap (up to 6.2 eV)[1-3], strong $\chi^{(2)}$ nonlinearity[4-6], high thermal conductivity and significant piezoelectric effect[7-9], which are typically not accessible in conventional silicon-based materials such as silica and silicon nitride. Leveraging these properties, AlN gained lots of interest for applications like second harmonic (SHG) generation[10-13], Pockels-effect-based modulators[14] and photodetectors in the ultraviolet (UV) and near-infrared (NIR) regimes[15-17]. In addition, as a member of the III-nitride-family materials, AlN shares a similar refractive index, nonlinear index, and strong $\chi^{(3)}$ nonlinearity with $Si_3N_4$[3,18]. As a result, integrated photonics devices based on AlN can be extremely versatile and robust for many integrated photonic applications[19-31].

In the past decades, different fabricating methods for high-quality ($Q$) factor AlN microresonators have been reported. However, all these methods require multiple layers of hard masks combined with a metal mask[32-36]. This complexity arises from the extremely high hardness of the material and thus low etching selectivity between AlN and photoresist, which requires multiple layers of hard masks to attain the desired etching depth. Additionally, strong charging of the material from electron beam (e-beam) lithography requires a conductive material (metal mask, typically Ti/Cr/Au) to avoid stitching errors and misalignment issues.

In this work, we report a novel fabrication process that utilizes a single hard mask layer without requiring any metal mask on top. To address the charging problem associated with e-beam lithography, we use a water-solvable conductive layer, which can be easily removed during the development step, simplifying the overall fabrication process. Our devices are fabricated using single-crystalline AlN on sapphire[36]. The fabricated AlN microresonators have a high intrinsic $Q$ factor of around $1.0 \times 10^6$ in the telecom band, enabling the successful generation of Kerr frequency combs around 1550 nm. Additionally, other nonlinear phenomena, such as Raman lasing, third harmonic generation (THG), and supercontinuum generation are also observed, which further verifies the suitability of the fabricated devices for nonlinear optics applications.

# II. Device fabrication and characterization

To fabricate the low-loss AlN nanophotonic devices, we start with an AlN-on-sapphire wafer with a single crystalline 1-µm-thick layer of AlN. To achieve good mode confinement and anomalous dispersion in the telecom band, we first determine the desired etching depth by simulating the mode profile, as shown in Fig. 1(a). We find that around 800 nm etching depth is enough for light confinement and achieving anomalous dispersion. Next, we evaluate the wafer surface roughness with an atomic force microscope (AFM), yielding a root mean square roughness of approximately 0.34 nm, as depicted in Fig. 1(b). The fabrication flow for realizing an AlN microresonator is illustrated in Fig. 1(c). We begin with reactive magnetron sputtering of a 250-nm-thick $Si_3N_4$ hard mask onto the AlN layer at room temperature[37]. This thickness of the hard mask is sufficient to survive ~800 nm AlN etching. To pattern the photonic structures, a layer of negative photoresist (ma-N 2405) is spin coated onto the $Si_3N_4$. In order to avoid charging effects, we spin-coat a conductive layer (mr-Conductive) on top of the photoresist. This water-solvable layer is later removed together with the photoresist, during the development step using developer ma-D 532, without requiring additional treatment. The pattern is then transferred to the $Si_3N_4$ mask by inductively coupled plasma reactive ion etching (ICP-RIE), using a gas mixture of $CHF_3$ and $O_2$. The AlN layer is then etched using a $Cl_2/BCl_3/Ar$ gas mixture at a ratio of 25:6:9, achieving an etching depth of around 800 nm. The etching selectivity between AlN and $Si_3N_4$ is approximately 4:1, which is sufficient for our process. The $Si_3N_4$ mask is almost completely removed during the AlN etching process, leaving only a very thin layer (tens of nm) of $Si_3N_4$ on top of the AlN. Given the similar refractive indices of $Si_3N_4$ and AlN, this thin $Si_3N_4$ layer does not disturb the mode profile and can be kept on the photonic structures. Finally, the sample undergoes cleaning with Piranha solution to remove any residual photoresist. A scanning electron microscope (SEM) image of the fabricated structure after AlN etching, shown in Fig. 1(d), reveals very small sidewall roughness.

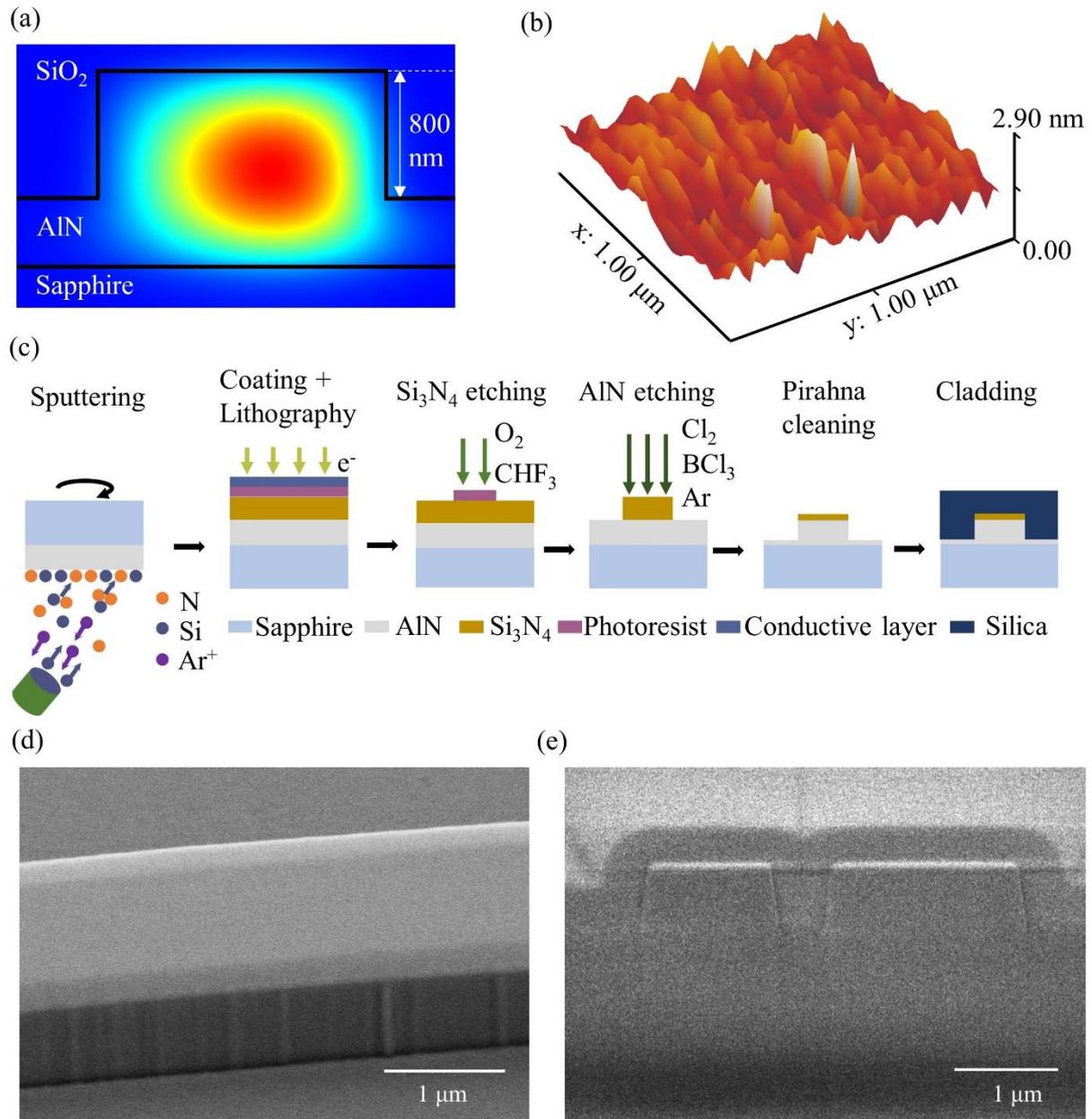

**Figure 1**. Metal-mask free AlN microresonator fabrication. (a) Simulation result of the mode profile for a partially-etched ring microresonator (~800 nm etching depth). (b) AFM image of a single crystalline AlN-on-sapphire sample with a 250-nm-thick $Si_3N_4$ mask on top. (c) Fabrication flow for the AlN process. (d) SEM image of an AlN waveguide. (e) FIB-SEM cross section of the coupling region between ring and waveguide.

To protect the photonic structures, a ~3-μm-thick $SiO_2$ cladding layer is deposited onto the chip. We perform a two-step silica deposition to avoid air voids between bus waveguide and resonator waveguide. The first 400 nm oxide is deposited using atomic layer deposition (ALD) and then another layer of 2.5 μm silica is deposited by plasma enhanced chemical vapor deposition. A Focused ion beam (FIB) SEM image of the coupling region between waveguide and ring resonator is shown in Fig. 1(e), where we can confirm that the gap between the ring and waveguide is filled properly by the ALD silica. Additionally, we can identify the AlN etching angle of ~81°. No annealing is performed after encapsulation. Finally, the sample is diced for edge coupling from a fiber to the bus waveguide. It should be noted that the sapphire substrate is extremely hard, making cutting the whole chip challenging and resulting in chipped edges. Instead, we partially saw cut into the sapphire side, followed by manually breaking the chip.

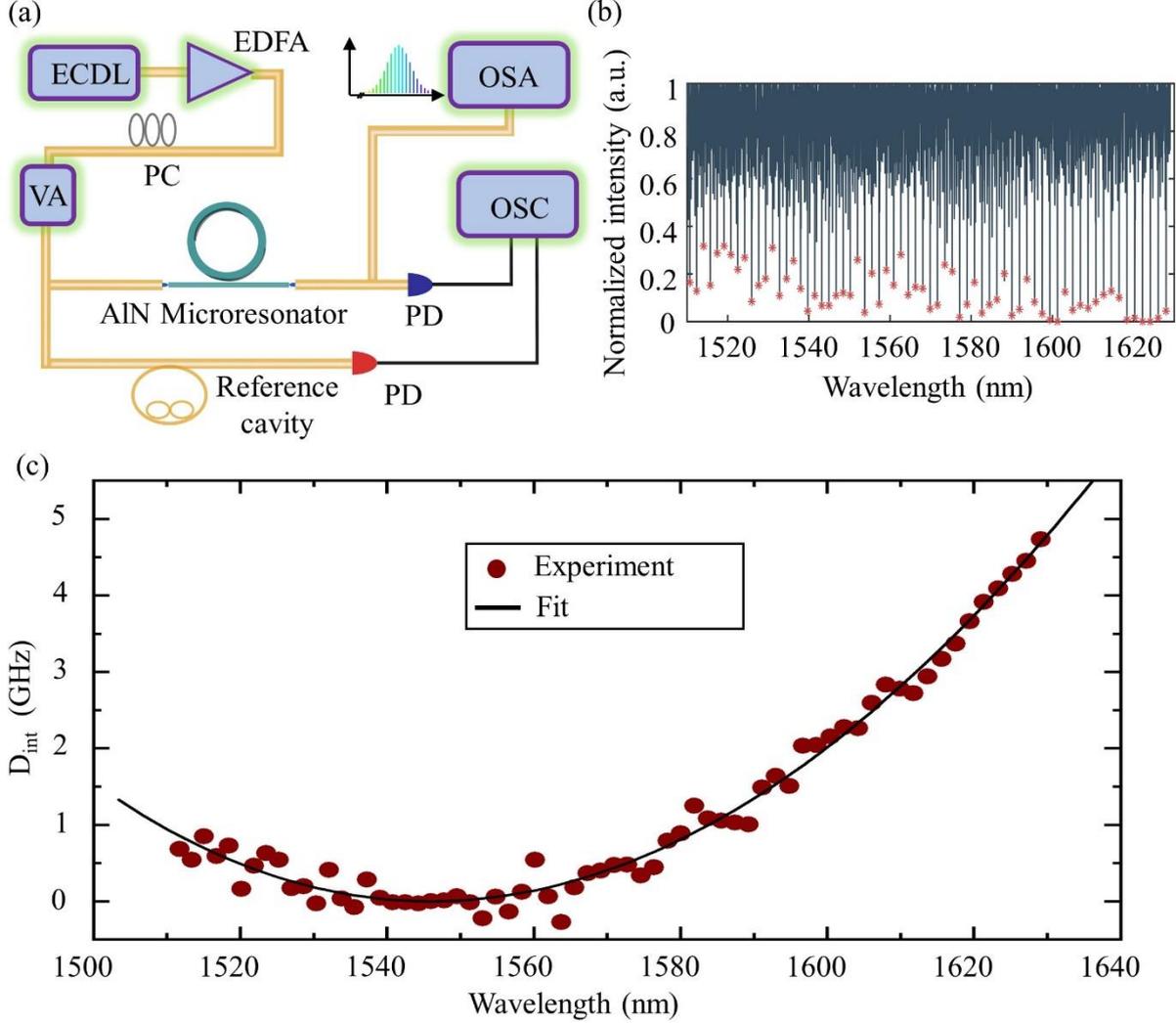

**Figure 2**. Experimental setup and device characterization. (a) Setup for characterization of the photonic chip and generating different types of nonlinear processes in AlN devices. ECDL: External cavity diode laser. EDFA: Erbium-doped fiber amplifier. PC: Polarization controller. VA: Variable attenuator. PD: Photodiode. OSC: Oscilloscope. OSA: Optical spectrum analyzer. (b) Resonance spectrum across the C-band and L-band. (c) Measured dispersion profile at the pump wavelength of 1546 nm and a corresponding second-order polynomial fit.

The fabricated devices are characterized with high-precision tunable diode laser spectroscopy[38] using the setup shown in Fig. 2(a). We use an external cavity diode laser (ECDL) in the C- and L-band to probe the AlN devices. Light is coupled in and out of the chip via edge coupling with two lensed fibers. The light polarization is controlled by a polarization controller (PC) to match either the quasi-transverse magnetic (TM) or the quasi-transverse electric (TE) modes of the waveguides. The measured total insertion loss from fiber to fiber is around 8 dB. The laser power in these measurements is sufficiently low to avoid Kerr and thermal effects. During frequency scanning, the laser frequency can be precisely determined by a fiber reference cavity calibrated by dual radio frequency modulation[38]. The transmission spectra of the AlN resonators and the reference cavity are recorded simultaneously. Fig 2(b) shows the transmission spectrum of the AlN ring resonator with 100-micron radius and 1.8-micron waveguide width. Using the measured resonance frequencies, we can calculate the dispersion profile of the device. The resonance frequencies are described as a function of the integrated dispersion[39] from the Taylor expansion:

$$\omega_k = \omega_0 + D_1 k + D_{int}(k) .$$

In the above formula, $\omega_k$ are the resonance frequencies with mode number $k$ (with $k = 0$ being an arbitrarily chosen center mode). $D_1$ indicates the free spectral range (FSR) at the center mode such that $D_1 = 2\pi \times$ FSR, and $D_{int}$ represents the integrated dispersion. The captured dispersion and the polynomial fits are shown as dark-red circles and black lines respectively in Fig. 2(c). Our sample exhibits anomalous dispersion, which is important for the generation of bright solitons.

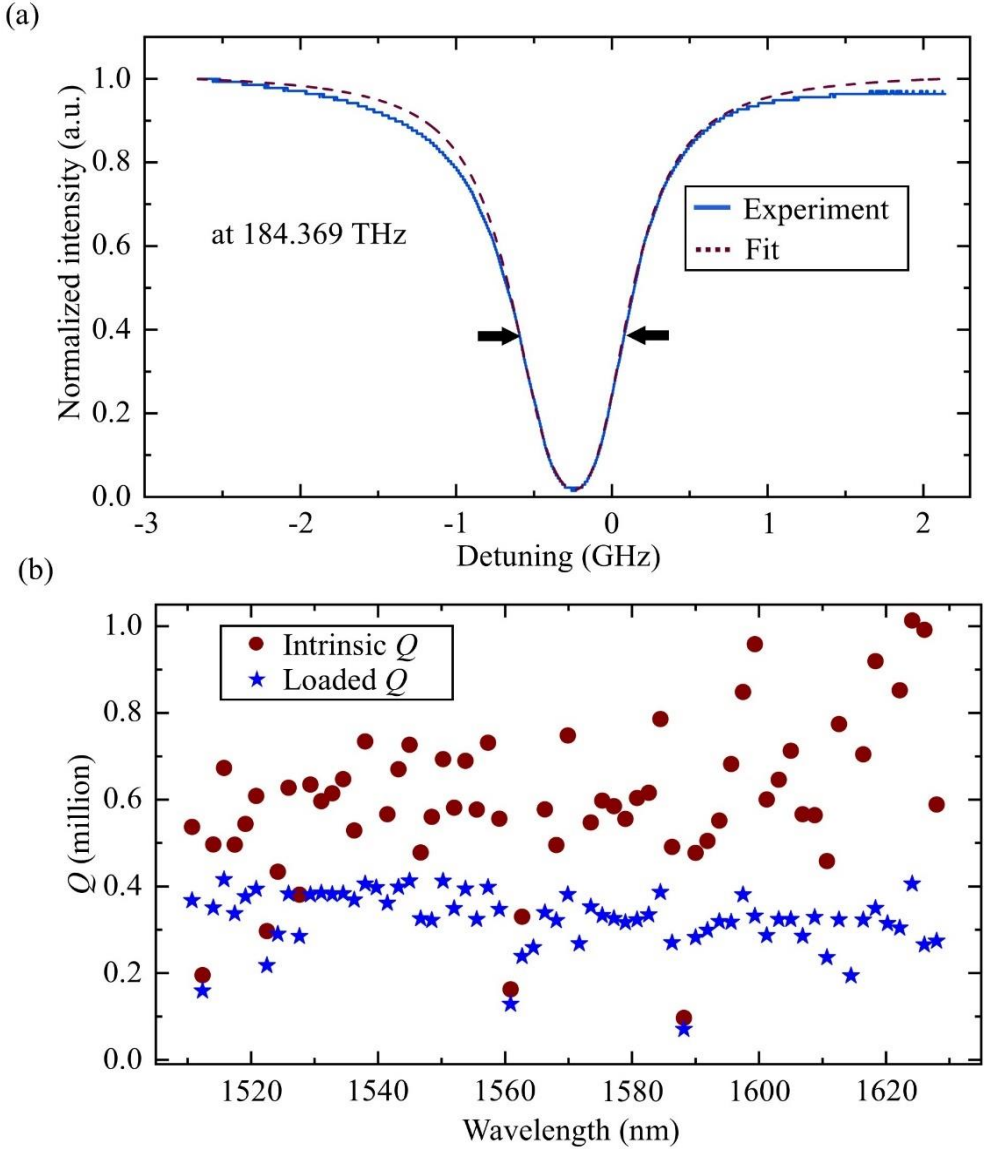

**Figure 3**. $Q$ factor measurements of the fabricated microresonators. (a) Normalized transmission spectrum for a high $Q$ mode. (b) Summarized loaded $Q$ factors and their corresponding intrinsic $Q$ factors across the C-band and L-band.

The $Q$ factors of the resonators are shown in Fig. 3. Fig. 3(a) shows a selected high $Q$ resonance around 1626 nm. The fitted linewidth of the resonance is around 694 MHz, with a loaded $Q$ of 0.27 million. The intrinsic $Q$ is calculated to be approximately 1.0 million, considering slight over-coupling of the bus waveguide to the resonator. Using the intrinsic $Q$ value and a free spectral range (FSR) of around 221 GHz, the propagation loss can be determined using the formula: $l = f_r / (R \times FSR \times Q_i)$, where $f_r$ indicates the resonance frequency and R is the ring resonator radius. Based on these parameters, the propagation loss is calculated to be 0.36 dB/cm. Fig. 3(b) shows a measurement of all the $Q$ factors of the resonances in the range between 1510 nm and 1630 nm. The average intrinsic $Q$ is 0.65 million.

## III. Nonlinear optics applications

In this section, we investigate the low loss AlN platform for nonlinear optical effects, including Kerr frequency comb generation, Raman lasing, THG and supercontinuum generation. These phenomena hold significant potential for various applications with integrated photonic circuits based on AlN.

## Kerr Frequency comb generation

Fig. 4(a), shows degenerate and non-degenerate four-wave mixing (FWM) processes resulting in the generation of a frequency comb spectrum. In the experiment, we pump the resonator with 800 mW on-chip power and tune the pump laser frequency into a resonance from the blue detuned side. The spectrum of the Kerr frequency comb is recorded by an optical spectrum analyzed (OSA) and shown in Fig. 4(b), where the pump laser is marked with a green cross. The comb lines span around 250 nm at an FSR of 221 GHz.

## Raman lasing

A simplified Raman lasing scheme is illustrated in Fig. 4(c), where the incident pump photon interacts with a phonon through inelastic scattering. More specifically, the incident photon can transfer its energy to the vibrational mode, and thus the emitted photon has lower energy and longer wavelength (Stokes shift). Alternatively, the incident photon can also gain energy from the excited vibrational mode, and thus the resulting emission has higher energy and shorter wavelength (anti-Stokes shift). Since the number of molecules in the vibrational ground state have a much bigger population than the excited ones, the power of the Stokes sideband is much stronger than the power in the anti-Stokes sideband. The wurtzite structure of AlN supports six different types of Raman phonons. In the experiment, we first align the polarization of the pump laser to the TM mode that is aligned with the AlN c-axis, then tune the laser frequency from the blue-detuned side into the resonance. When the intra-cavity power gradually increases, we observe a Raman Stokes signal at 1715.3 nm and an anti-Stokes signal at 1419.2 nm as shown in Fig. 4(d). The Raman shift with respect to the pump laser is ~18.2 THz. This shift corresponds to an $A_1^{TO}$ phonon[31].

## Third harmonic generation

The strong $\chi^3$ nonlinearity of AlN supports third harmonic generation, as depicted in Fig. 4(e). By pumping a mode at 1538.2 nm, we observe the emission of strong green light. The conversion of light from telecom regime to the infrared regime is captured by a visible camera as shown in Fig. 4(f). The phase matching of the THG process happens at fixed detuning values of the pump laser with respect to the resonator mode and is determined by the resonator dispersion.

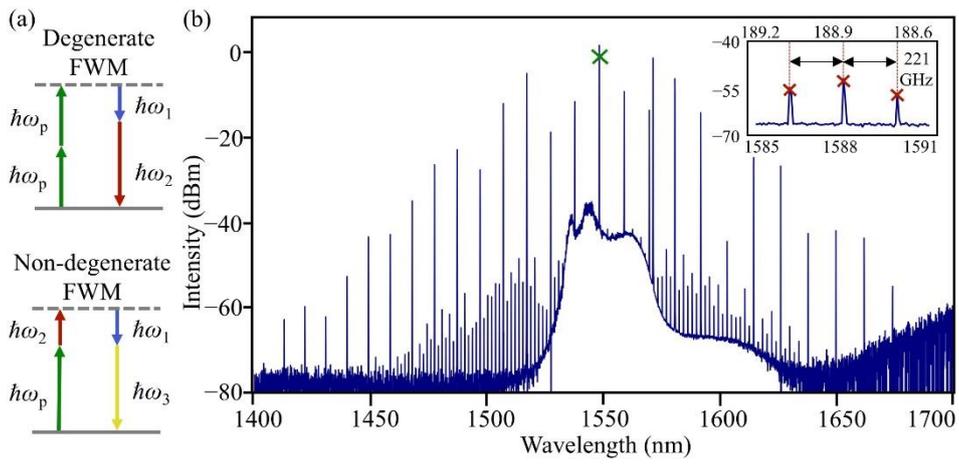

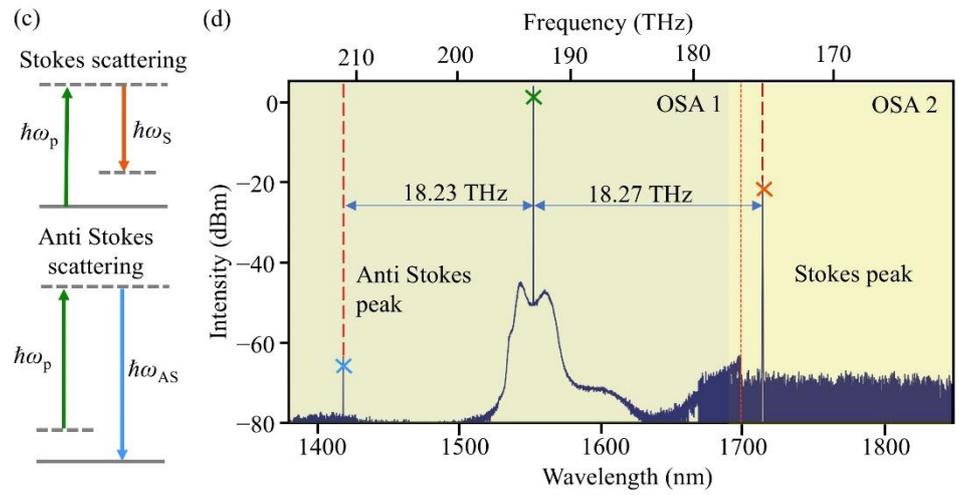

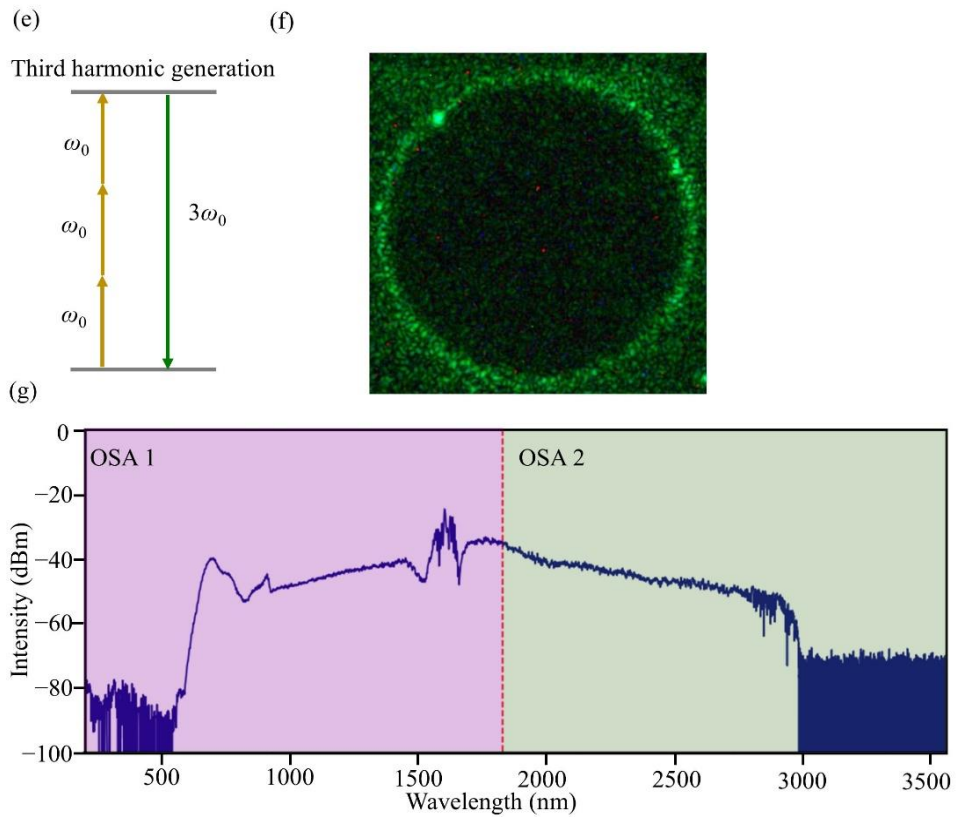

**Figure 4**. Various nonlinear effects in AlN. (a) Energy diagram for frequency comb generation through four-wave mixing. (b) Measured comb spectrum. (c) Energy diagram for Raman scattering. (d) Measured Stokes and anti-Stokes sidebands. (e) Energy diagram for third-harmonic generation. (f) Third harmonic light emission from ring resonator. (g) Supercontinuum generation covering ~2.5 octaves from 600 nm to 2.9 μm.

## Supercontinuum generation

Finally, we demonstrate supercontinuum generation in an AlN waveguide, which is a mediated by a combination of nonlinear phenomena, including FWM, self-phase modulation, cross-phase modulation and harmonic generation. A 50-femtosecond pulse train from an Er-fiber frequency comb with 100 MHz repetition rate and 1560 nm central wavelength is used to pump the nonlinear waveguide. The pulses are coupled into the bus waveguide using an aspheric lens (0.6 numerical aperture) with an estimated insertion loss of -5 dB/facet. Driven by the femtosecond pulse excitation, the supercontinuum emerges as a broadband spectrum resulting from the interplay of nonlinear processes as well as optimized dispersion. We observe a two-octave spectrum from VIS to Mid-IR, which is collected by an InF3 fiber, shown in Fig. 4(f). The average on-chip power is around 10 mW. The large spectral broadening takes advantages of both the large Kerr nonlinearity and high confinement of the AlN waveguide and has significant potential for applications in optical frequency metrology, high-resolution spectroscopy and optical telecommunications.

## IV. Conclusion

In conclusion, we demonstrate a simple fabrication process for low-loss AlN-based nanophotonics without the requirement for metal masks and thermal annealing. With this process we achieve *Q*-factors of 1.0 million in AlN microresonators with a corresponding propagation loss of 0.36 dB/cm. We demonstrate Kerr frequency combs, Raman lasing, THG and supercontinuum generation, highlighting the suitability of this platform for a wide range of applications.

## Acknowledgement

This work was supported by the Deutsche Forschungsgemeinschaft project 541267874, European Union's H2020 ERC Starting Grant 756966, the Marie Curie Innovative Training Network "Microcombs" 812818, and the Max Planck Society. We thank Anette Daurer from Fraunhofer IISB for help with chip dicing. We thank the Technology Development and Service Group for Nanofabrication at MPL for support with the cleanroom processes.